 \documentstyle[preprint,aps]{revtex}

\begin{document}

 \draft

 \title{A model for the shapes of islands and pits on\\
(111) surfaces of fcc metals}
 \author{G. T. Barkema and M. E. J. Newman}
 \address{Laboratory of Atomic and Solid State Physics, Cornell University\\
Ithaca, NY 14853--2501.}
 \author{M. Breeman}
 \address{Nuclear Solid State Physics, Materials Science Centre,
Groningen University\\
Nijenborgh 4, 9747 AG Groningen, The Netherlands.}

 \maketitle

 \begin{abstract} It is experimentally observed that adsorbate atoms and
vacancies on (111) surfaces of fcc metals cluster into islands which are
approximately hexagonal, but which on closer inspection turn out to have
equilibrium facets which alternate in length $ABABAB$ around the six sides
of the island.  By contrast, previous theoretical models for island faceting
predict a rotating sequence of three lengths $ABCABC$ around the island.  We
propose a new model for the observed shapes, whose physical basis is the
variation of the local arrangements of substrate atoms seen by an adsorbate
atom.  We map our model onto a generalized form of the two-dimensional Ising
model having three- as well as two-spin interactions, and estimate using
atom-embedding calculations the strengths of these interactions for Cu
adsorbed on a Cu~(111) surface.  We then describe a new, highly efficient
Monte Carlo technique for calculating the equilibrium crystal shapes of
general Ising-type models in two or three dimensions, and apply it to the
model in hand.  Our results do indeed show alternating facet lengths closely
similar to those seen in experiments.
\end{abstract}

 \narrowtext

 \section{Introduction} With the introduction of the technique of scanning
tunneling microscopy (STM) it has become possible to study overlayer
structures on surfaces in real space with resolution on the scale of single
atoms.  Of the many interesting surface features this technique has
revealed, the ones that will occupy us in the present work are islands of
adsorbate atoms on (111) surfaces of metals, and in particular the shapes of
these islands, which can be observed either under growth
conditions~\cite{bott}, or in thermal
equilibrium~\cite{michely1,michely2,michely3}.  During growth, and
especially at low temperatures, island shapes can be influenced by the
kinetics of the growth process.  However, at higher temperatures, kinetic
effects are of less importance, and island shapes are governed principally by
thermodynamics.  There is some experimental evidence that such an
equilibrated high temperature regime does exist.  In STM experiments on the
growth of Pt on Pt~(111), Michely and co-workers~\cite{michely2} have
compared the shapes of islands formed following sputtering at high
temperatures (around 700K) and following high temperature annealing of
surfaces sputtered at lower temperatures.  Apart from transient effects
which are well explained by the details of the experimental technique,
the islands are qualitatively very similar, implying that the higher
temperatures fall in a regime in which thermal equilibration of a lattice
gas of adatoms or vacancies is the dominant effect on island morphology, as
opposed to kinetic effects of the sputtering.  Another indicative result is
that, in this high temperature regime, the islands have a very regular
shape, varying little from one island to another, or from large islands to
small ones.  This behavior is typical of a lattice gas in an equilibrium or
near-equilibrium state.

For a thermally equilibrated system, the concept of `equilibrium crystal
shape' (ECS) of adatom islands on surfaces may provide insight into the
character of the interactions between adatoms and the interactions between
adatoms and the substrate.  Since the development of atom-embedding methods
like the EAM method of Daw and Baskes~\cite{daw}, the Effective Medium
Theory~\cite{hansen}, and the Finnis-Sinclair method~\cite{finnis}, the
calculation of the energetics of larger adatom clusters on surfaces has
become feasible.  In a recent paper~\cite{breeman}, we have studied the
binding energies and stability of Cu-adatom islands on Cu~(100) and Cu~(111)
surfaces by means of the atom-embedding method of Finnis and Sinclair.  For
islands on Cu~(111) it was found that the binding energies of adatom clusters
are dominated by nearest-neighbor interactions, with next-nearest-neighbor
interactions playing only a minor role.  If the binding energy of a cluster
were simply proportional to the number of nearest-neighbor-bonds in the
cluster, hexagonal island shapes having all sides of equal length would be
preferred.  However, the experiments of
Michely~{\it{}et~al.}~\cite{michely2} show hexagonal islands with facets
alternating in length around the island.  As the experimenters have pointed
out, this observation can be understood by noting that there are two possible
step-edge (and thus island edge) orientations on a (111) surface of an fcc
material; steps either have a \{100\}-oriented edge or a \{111\}-oriented
edge.  These two types of edge alternate around the perimeter of a hexagonal
surface island, and if the two were equally energetically favorable we would
expect the island to be perfectly hexagonal.  The fact that the experiments
show alternating unequal edge lengths implies that actually one orientation
is energetically favored over the other.  In our previous
study~\cite{breeman}, the \{100\}-oriented step was found to have a lower
energy for Cu islands on Cu~(111), in agreement with experiment~\cite{klas}.
Experimentally, the reverse seems to be true for Pt~\cite{michely2}, for
which the \{111\}-oriented step is the energetically favored one, giving
rise to surface islands with the opposite orientation to those observed for
Cu.  In this paper we calculate the ECS of adatom islands on the (111)
surface of an fcc metal by means of Monte Carlo simulations of a
two-dimensional triangular-lattice Ising-type model.  We use a Hamiltonian
whose form reflects the physical origin of the different line tensions for
the two facet types, and the parameters appearing in the Hamiltonian are
fitted to the results of atom-embedding calculations for Cu islands on
Cu~(111)~\cite{breeman}.  Although our calculations are all for islands of
adatoms, the model is equally well suited to the calculation of the
equilibrium shapes of pits in an overlayer of adatoms on a (111) surface.
In fact, because of the asymmetric form of the line tension within our
model, we expect the ECS for a surface pit to be exactly the same as that
for an island of adatoms, for sufficiently large islands, except that it
will be rotated by $180^\circ$.  It is interesting to note that the
experiments of Michely and co-workers display precisely this
phenomenon~\cite{michely2}.

In Section~II we describe our model.  In Section~III we give details of the
new Monte Carlo technique which we have developed to calculate ECS's for
models of this type, and in Section~IV we describe the results of our
simulations, which show hexagonal surface islands with facets alternating in
length $ABABAB$ around their perimeter, in agreement with experiment.  In
Section~V we give our conclusions.

 \section{The model} Following a number of previous studies (for example,
Ref.~\cite{zia}), we use a form of the Ising model on a triangular lattice in
which spins of one sign (say $s=+1$ or `up' spins) represent sites on the
surface which are occupied by adsorbate atoms, and spins of the opposing
sign ($s=-1$ or `down' spins in this case) represent sites which are
unoccupied.  Surface islands are then represented by domains of up spins in
a sea of down spins.  Surface pits are represented by domains of down spins
in a sea of up.  The Hamiltonian includes nearest-neighbor ferromagnetic
interactions to encourage the aggregation of islands and pits, and we want
to add extra terms to make the model duplicate the alternating long and
short facet behavior seen in experimental observations.  This necessarily
means that opposite facets on the same island should have different
equilibrium lengths, even though they run in the same direction, which in
turn tells us that two-spin interactions will be inadequate to reproduce
this behavior in our Ising system; these opposite sides will always have the
same line tension and thus the same equilibrium length when only two-spin
interactions are present.  This is seen clearly in the work of
Zia~\cite{zia} who studied ECS's for Ising models with the most general form
of nearest-neighbor two-spin anisotropic interactions on a triangular
lattice.  The general form of the ECS for these models has three different
equilibrium facet lengths rotating in the sequence $ABCABC$ around the
hexagonal island.  In all cases, opposite sides have the same length.

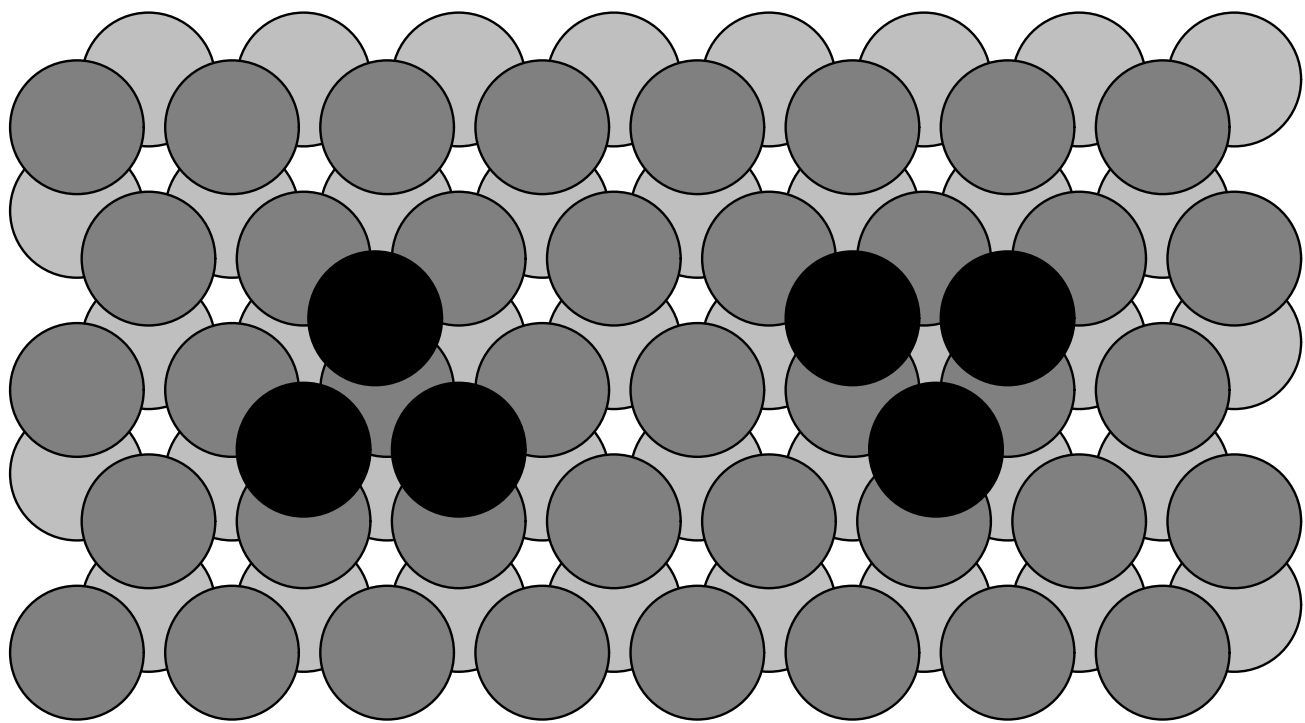
 \begin{figure}
 \caption{Up-pointing and down-pointing triangles of adatoms on a (111)
surface.  Notice that the up-pointing triangle has a substrate atom
immediately beneath it, whereas the down-pointing one does not.  Because of
this, interactions between the adatoms and the substrate give rise to a
difference between the energies of these two configurations.}
 \label{triangles}
 \end{figure}

In order to understand how the model should be generalized to produce the
behavior seen in the experimental systems, we consider the physical mechanism
which accounts for the alternating equilibrium facet lengths.  This
mechanism is illustrated in Figure~\ref{triangles}.  We see that the
adsorbate atoms on a triangular lattice can form two different three-atom
triangles: the up-pointing ones $\bigtriangleup$ and the down-pointing ones $\bigtriangledown$.
The local environments of the two types of triangles differ because the
atoms of one type (in this case, the up-pointing ones) surround an atom in
the substrate, whereas those of the other type (down in the figure) do not.
This may either increase or decrease the contribution to the binding energy
of the up-pointing configuration over that of the down-pointing one,
depending on the details of the interaction between adsorbate and substrate
atoms.  In Cu, for instance, both our calculations and experiment
indicated that the up-pointing triangle in the figure should be favored over
the down-pointing one, whereas in Pt the situation appears to be reversed.

Within the bulk of an island this difference in energies has no effect,
since every adatom belongs to six triangles, three of which are
up-pointing and three down-pointing, making every atom energetically
equivalent to every other.  However, the atoms on the edges of the island
belong to only three triangles---either two up and one down, or {\it vice
versa,} depending on the orientation of the edge.  The imbalance in the
symmetry of the local environments between these two cases gives rise to a
difference in the line tensions of the two types of edges, and thus gives
two different equilibrium facet lengths for the island.

To reproduce this behavior in our model we need to introduce a three-spin
contribution to the Hamiltonian, which is non-zero only for those sets of
three nearest-neighbor sites around a triangle which are all occupied by an
adsorbate atom (i.e.,~for which all three spins are in the up state).  Such
an interaction term is of the form
 \begin{equation}
 \label{triangle}
 \sigma_i\sigma_j\sigma_k + \sigma_i\sigma_j + \sigma_j\sigma_k +
\sigma_k\sigma_i + \sigma_i + \sigma_j + \sigma_k + 1,
 \end{equation}
 where $\sigma_i$, $\sigma_j$, $\sigma_k$ are the three Ising spins around
the triangle.  We multiply this by some interaction energy which takes
one of two different values, $K^\bigtriangleup$ and $K^\bigtriangledown$, depending on whether
the triangle is oriented pointing upwards or downwards, and sum over all
trios of spins $\langle\sigma_i\sigma_j\sigma_k\rangle$ arranged in a
triangle.  Notice however that in doing this, each pair-wise interaction
$\sigma_i\sigma_j$ appears exactly twice, once multiplied by $K^\bigtriangleup$, and
once by $K^\bigtriangledown$.  Thus the two-spin terms in~(\ref{triangle}) can be
accounted for by simply adjusting the nearest-neighbor coupling already
appearing in the model according to $J \to J + K^\bigtriangleup + K^\bigtriangledown$.  The
terms linear in the spins can be rewritten as a simple sum over the lattice
$6\sum_i\sigma_i$.  We will be using conserved order-parameter (COP)
dynamics to investigate the formation of islands with diffusing adsorbates,
and with a conserved order parameter this linear term is just a constant.
Thus, to within a constant, our Hamiltonian takes the form:
 \begin{equation}
 \label{hamiltonian}
 H = - J \sum_{\langle i j \rangle} \sigma_i\sigma_j
     - K^\bigtriangleup \sum_{\langle i j k \rangle_\bigtriangleup}
	 \sigma_i\sigma_j\sigma_k
     - K^\bigtriangledown \sum_{\langle i j k \rangle_\bigtriangledown}
       \sigma_i\sigma_j\sigma_k.
 \end{equation}
 Here the notation $\langle i j k \rangle_\bigtriangleup$ means that the
three sites $i$, $j$, $k$ are nearest-neighbors arranged around a
triangle oriented upwards, and conversely for $\langle i j k
\rangle_\bigtriangledown$.

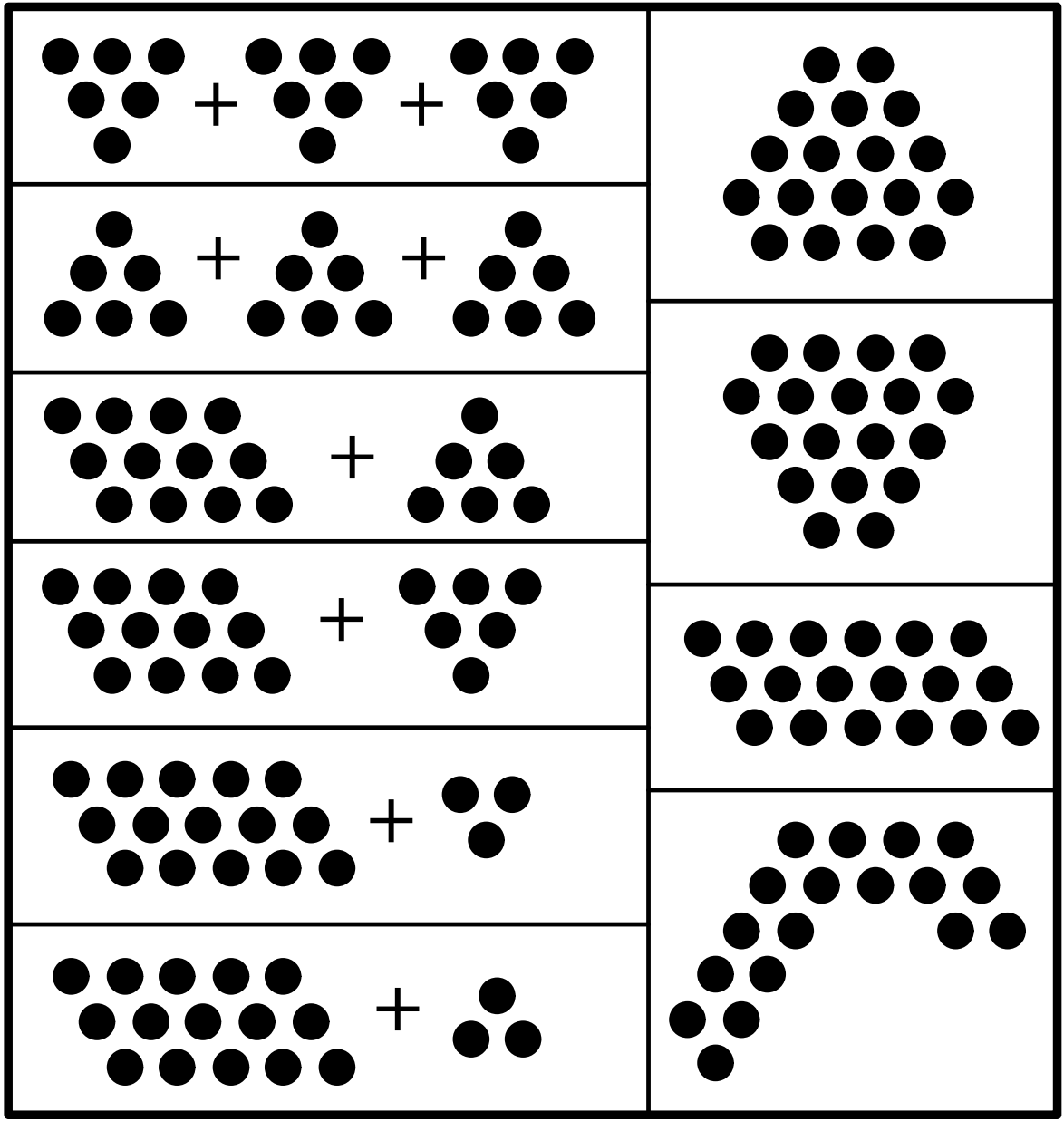
 \begin{figure}
 \caption{The ten different configurations of eighteen Cu adatoms on a Cu
(111) surface used to fit the parameters in the Ising Hamiltonian to
energies calculated using the atom-embedding method.}
 \label{islands}
 \end{figure}

Using the atom-embedding method of Finnis and Sinclair~\cite{finnis}, we have
calculated the binding energies of the ten different 18-atom adsorbate
islands shown in Figure~\ref{islands} for Cu adsorbed on Cu~(111).  We find
that our Hamiltonian fits the energies in these calculations to a very good
approximation (about $0.012{\rm eV}$) with $J = 0.144{\rm eV}$ and $K^\bigtriangleup
= -K^\bigtriangledown = 0.0063{\rm eV}$.  The effect of the underlying substrate atoms
is thus a small one by comparison with the nearest-neighbor coupling, and
can to a good approximation be ignored in the calculation of the critical
temperature for island formation, which is then given by the corresponding
expression for the ordinary Ising model on a triangular lattice: $\beta J =
\arctan(1 - {\rm e}^{\beta J})$~\cite{baxter}.  When we come to calculate the
equilibrium island shape however, the small three-spin term has an important
effect on the facet lengths.

 \section{The Monte Carlo algorithm} Our strategy for calculating the ECS
for our model is to place a certain number of atoms (i.e.~up spins) in an
island on the triangular lattice and then to allow them to diffuse under
some conserved order-parameter Monte Carlo dynamics.  We take the
resulting distribution of island shapes, superimpose them by identifying
their centers of mass, and then average over them to find the mean ECS.
The exact same method can also be used to calculate the ECS for a pit in the
surface.

The simplest COP dynamics is the Kawasaki spin flip, in which adjacent pairs
of oppositely-oriented spins are flipped with a probability that depends
exponentially on the difference in energies between the states of the model
before and after the flip.  For the present study however this would be an
extremely inefficient dynamics to choose, since the motion of atoms from one
part of the island to another is by diffusion in single-lattice-parameter
steps.  Furthermore, the high energetic cost of pulling an atom away from
the island edge means that most changes in the shape of the island would take
place by edge diffusion, making equilibration still slower.  However,
since we are not interested in the way in which we reach the ECS, but only
in the final result, there is no need to confine ourselves to Monte Carlo
moves which are locally order-parameter conserving, like the Kawasaki move,
and in fact it turns out that we can do much better by using a non-local
move.

The simplest non-local move would be one in which we randomly choose an atom
and a vacancy and interchange them, again with an acceptance probability that
depends exponentially on the difference in energy between the initial and
final states.  However, this is also inefficient, since most such moves will
involve taking an atom from the bulk of the island and depositing it at some
isolated position outside the island.  This procedure will have a high energy
cost making the acceptance ratio very low.  So instead we propose a new
algorithm in which, rather than choosing an atom and a vacancy completely at
random from the available possibilities, we choose some atoms and vacancies
with a higher probability than others, in just such a way, that if we always
make the resulting swap of atom and vacancy, detailed balance is exactly
obeyed and a Boltzmann equilibrium distribution of states results.  In order
to describe our algorithm, we need first to define some quantities.

The nearest-neighbor coordination number $N_i$ of an adatom on the $i^{\rm
th}$ site is the number of adjacent surface sites also occupied by adatoms:
 \begin{equation}
 N_i = \sum_{\langle ij\rangle} \delta_{\sigma_i\sigma_j}.
 \end{equation}
 For a vacant surface site, this formula gives the number of adjacent
vacancies to that site, and this turns out to be a useful generalization of
the coordination number concept, since $N_i$ is then the number of satisfied
(parallel) bonds to site $i$ in our Ising representation of the problem.  We
also define the up- and down-triangle coordination numbers $T^\bigtriangleup_i$ and
$T^\bigtriangledown_i$ for the $i^{\rm th}$ site which are the number of up and down
triangles respectively of which the site $i$ is a member and on which all
the spins are similarly oriented to the one on site $i$ (all occupied by
adsorbate atoms, or all vacant).  We can write $T^\bigtriangleup_i$ and $T^\bigtriangledown_i$
in the form:
 \begin{eqnarray}
 T^\bigtriangleup_i &=& \sum_{\langle i j k \rangle_\bigtriangleup}
\delta_{\sigma_i\sigma_j}\delta_{\sigma_i\sigma_k},\nonumber\\
 T^\bigtriangledown_i &=& \sum_{\langle i j k \rangle_\bigtriangledown}
\delta_{\sigma_i\sigma_j}\delta_{\sigma_i\sigma_k}.
 \end{eqnarray}
 Notice that we can equivalently write
 \begin{eqnarray}
 T^\bigtriangleup_i &=& \frac14\sum_{\langle i j k \rangle_\bigtriangleup}
(\sigma_i\sigma_j + \sigma_j\sigma_k + \sigma_k\sigma_i + 1),\nonumber\\
 T^\bigtriangledown_i &=& \frac14\sum_{\langle i j k \rangle_\bigtriangledown}
(\sigma_i\sigma_j + \sigma_j\sigma_k + \sigma_k\sigma_i + 1).
 \label{useful}
 \end{eqnarray}
 This becomes useful in the proof of our algorithm.

 It now turns out that the correct algorithm for equilibrating the system is
to choose an occupied site $i$ with probability $P_i$ proportional to
${\rm e}^{-2\beta J N_i-4\beta K^\bigtriangleup T^\bigtriangleup_i-4\beta K^\bigtriangledown T^\bigtriangledown_i}$, and a
vacant site $j$ with probability $Q_j$ proportional to ${\rm e}^{-2\beta
JN_j+4\beta K^\bigtriangleup T^\bigtriangleup_j+4\beta K^\bigtriangledown T^\bigtriangledown_j}$.  The correct
normalizing factors for the probabilities are simply the sums of these
expressions over all occupied and unoccupied sites respectively:
 \begin{eqnarray}
 P_i &=&
 {{\rm e}^{-2\beta J N_i-4\beta K^\bigtriangleup T^\bigtriangleup_i-4\beta K^\bigtriangledown T^\bigtriangledown_i}
 \over\sum_{i|\sigma_i=+1}
 {\rm e}^{-2\beta J N_i-4\beta K^\bigtriangleup T^\bigtriangleup_i-4\beta K^\bigtriangledown T^\bigtriangledown_i}},
 \nonumber\\
 Q_j &=&
 {{\rm e}^{-2\beta J N_j+4\beta K^\bigtriangleup T^\bigtriangleup_j+4\beta K^\bigtriangledown T^\bigtriangledown_j}
 \over\sum_{j|\sigma_j=-1}
 {\rm e}^{-2\beta J N_j+4\beta K^\bigtriangleup T^\bigtriangleup_j+4\beta K^\bigtriangledown T^\bigtriangledown_j}}.
 \end{eqnarray}

If we swap two sites chosen in this fashion then detailed balance is {\em
almost\/} preserved.  The only reason why it isn't is that the probability
per unit time of making a particular move now depends on what other moves
are possible; a move taking an atom from one part of the island's boundary
to another, that would be a reasonably probable move in the normal run of
things, suddenly becomes very unlikely if there is, for instance, a vacancy
in the middle of the island somewhere.  The very large binding energy of an
atom in this site makes a move that takes a surface atom and fills this
vacancy with it overwhelmingly the most likely move in such a case, provided
we are a reasonable way below the transition temperature.  In order to
ensure that the probability of a certain move being made per unit time does
not vary in this way, we adjust our time variable $t$ not by unity at each
time-step, but by an amount
 \begin{equation}
 {1\over\Delta t} = \sum_{i|\sigma_i=+1}
 {\rm e}^{-2\beta J N_i-4\beta K^\bigtriangleup T^\bigtriangleup_i-4\beta K^\bigtriangledown T^\bigtriangledown_i}
 		    \sum_{j|\sigma_j=-1}
 {\rm e}^{-2\beta J N_j+4\beta K^\bigtriangleup T^\bigtriangleup_j+4\beta K^\bigtriangledown T^\bigtriangledown_j}.
 \label{timestep}
 \end{equation}

This algorithms fulfills the conditions of detailed balance and ergodicity,
and therefore samples the Boltzmann distribution corresponding to the
Hamiltonian~(\ref{hamiltonian}).  Ergodicity is satisfied, since it is
possible ultimately to reach any state, within the COP constraint, by
exchanging atoms with vacancies on the lattice.  To see that detailed
balance is satisfied, we write the rate $T_{ij}$ for a step which exchanges
an atom at site $i$ with a vacancy at site $j$ as
 \begin{equation}
 T_{ij} = {P_i Q_j\over\Delta t} =
 {\rm e}^{-2\beta J N_i-4\beta K^\bigtriangleup T^\bigtriangleup_i-4\beta K^\bigtriangledown T^\bigtriangledown_i}
 {\rm e}^{-2\beta J N_j+4\beta K^\bigtriangleup T^\bigtriangleup_j+4\beta K^\bigtriangledown T^\bigtriangledown_j},
 \end{equation}
 and the probability of the reverse move as
 \begin{equation}
 T_{ij} = {P_j' Q_i'\over\Delta t'} =
 {\rm e}^{-2\beta J N_j'-4\beta K^\bigtriangleup {T^\bigtriangleup_j}'-4\beta K^\bigtriangledown {T^\bigtriangledown_j}'}
 {\rm e}^{-2\beta J N_i'+4\beta K^\bigtriangleup {T^\bigtriangleup_i}'+4\beta K^\bigtriangledown {T^\bigtriangledown_i}'},
 \end{equation}
 where the primed variables indicate values of the various quantities in the
second state.  Then the ratio of the rates is
 \begin{equation}
 {T_{ij}\over T_{ji}} = {\rm e}^{-\beta \Delta E},
 \end{equation}
 where
 \begin{eqnarray}
 \Delta E &=& 2 J (N_i - N_i') + 2 J (N_j - N_j') -\nonumber\\
	  & & \quad 4 K^\bigtriangleup (T^\bigtriangleup_i + {T^\bigtriangleup_i}') -
		    4 K^\bigtriangledown (T^\bigtriangledown_i + {T^\bigtriangledown_i}') +
		    4 K^\bigtriangleup (T^\bigtriangleup_j + {T^\bigtriangleup_j}') +
		    4 K^\bigtriangledown (T^\bigtriangledown_j + {T^\bigtriangledown_j}').
 \label{deltae}
 \end{eqnarray}
 The terms in $N_i - N_i'$ and $N_j - N_j'$ give the change in energy
between the two states, due to the change in the number of satisfied bonds.
The terms like $T^\bigtriangleup_i + {T^\bigtriangleup_i}'$ give the change in energy due to
the triangle interactions.  To see this we note that site $i$ is occupied in
the first state ($\sigma_i=+1$) and vacant in the second state
($\sigma_i'=-1$) and then make use of Equation~(\ref{useful}) to write
 \begin{eqnarray}
 T^\bigtriangleup_i + {T^\bigtriangleup_i}' &=& \sigma_i T^\bigtriangleup_i - \sigma_i'
{T^\bigtriangleup_i}'\nonumber\\
			  &=& \frac14\sigma_i\sum_{\langle ijk\rangle_\bigtriangleup}
 (\sigma_i\sigma_j + \sigma_j\sigma_k + \sigma_k\sigma_i + 1) -
			      \frac14\sigma_i'\sum_{\langle ijk\rangle_\bigtriangleup}
 (\sigma_i'\sigma_j' + \sigma_j'\sigma_k' + \sigma_k'\sigma_i' + 1)
 \nonumber\\
			  &=& \frac32 + \frac14\sum_{\langle ijk\rangle_\bigtriangleup}
 (\sigma_i\sigma_j\sigma_k - \sigma_i'\sigma_j'\sigma_k').
 \end{eqnarray}
 In a similar way, we can rewrite the $T^\bigtriangleup_j + {T^\bigtriangleup_j}'$ term
in~(\ref{deltae}) as
 \begin{equation}
 T^\bigtriangleup_j + {T^\bigtriangleup_j}' = \frac32 - \frac14\sum_{\langle jkl\rangle_\bigtriangleup}
 (\sigma_j\sigma_k\sigma_l - \sigma_j'\sigma_k'\sigma_l').
 \end{equation}
 When we take $4K^\bigtriangleup$ times the difference of these two expressions, the
constant terms cancel, and we are left with precisely the change in the
contribution to the energy from the up-triangles.  A similar calculation
shows the same to be true for the down-triangles, and thus detailed balance
is obeyed.

This type of Monte Carlo algorithm is more general than the particular model
to which we have applied it in this paper.  Similar algorithms can be used
to rapidly calculate equilibrium faceting shapes in any conserved
order-parameter Ising-type model, and in particular we are not constrained to
surface islands; we could equally well use this type of algorithm to
calculate three-dimensional crystal shapes.  Its use in the calculation of
the ECS for simple-cubic and face-centered-cubic crystal shapes with
next-nearest-neighbor interactions in three-dimensions has been investigated
by Barkema and Holzer~\cite{barkema}.

 \section{Results} 
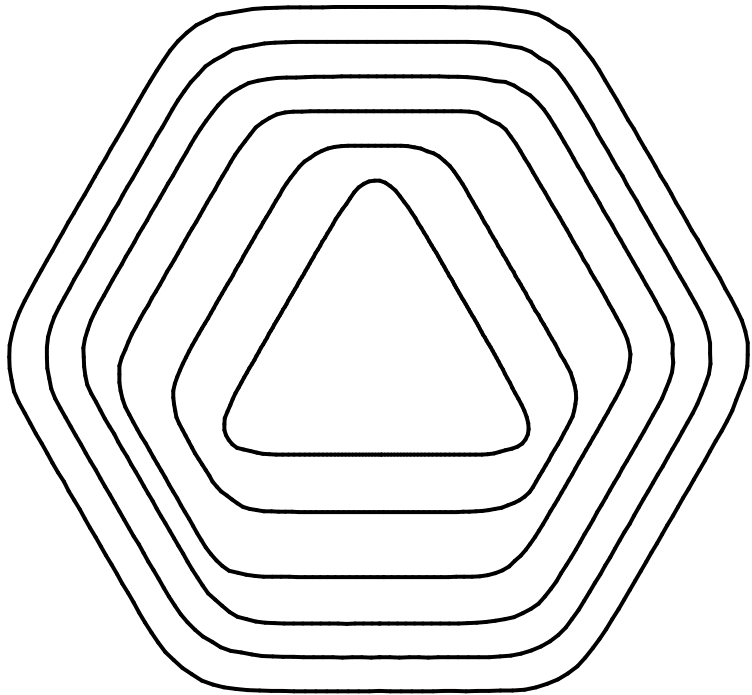
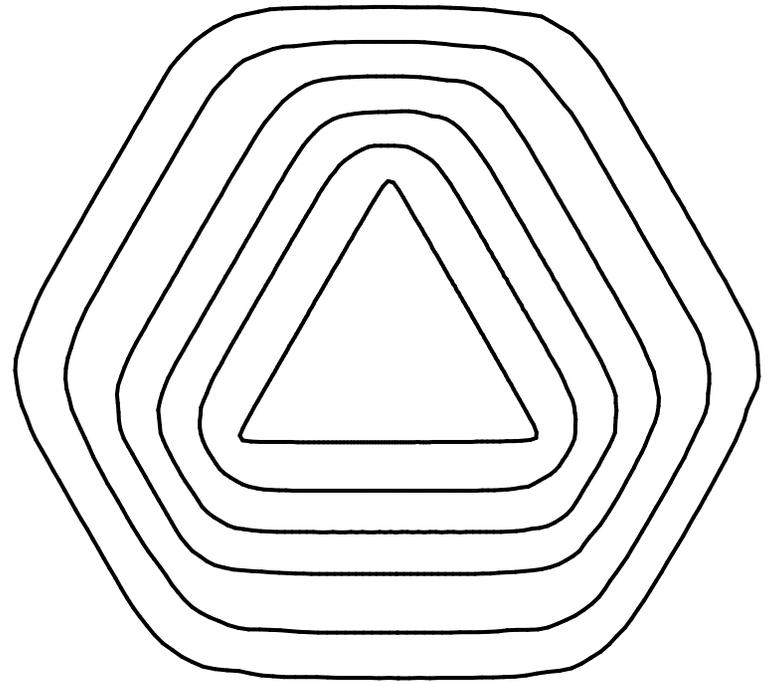
 \begin{figure}
 \caption{The equilibrium faceting shapes of islands of Cu adatoms on a Cu
(111) surface calculated within the model described in Section~II, using the
Monte Carlo method of Section~III.  The different curves are for different
values of the parameter $K$, increasing from outside to inside in the
sequence $0.005{\rm eV}$, $0.01{\rm eV}$, $0.02{\rm eV}$, $0.03{\rm eV}$,
$0.05{\rm eV}$, and for the innermost curve $0.07{\rm eV}$ on the left and
$0.1{\rm eV}$ on the right.  The curves on the left are the results at
$T=300$K and on the right at $T=600$K.  The shapes represent the lines of 50
percent occupation of sites averaged over $\sim10^7$ Monte Carlo steps, as
described in Section~IV.}
 \label{results}
 \end{figure}

We have used our Monte Carlo algorithm to calculate
equilibrium crystal shapes for islands in our model.  The results are shown
in Figure~\ref{results}, for islands consisting of 1875 adatoms on a lattice
of $7500$ sites.

Each run takes about ten minutes of CPU time on an IBM RS/6000 workstation.
The exact number of Monte Carlo steps used varies from one run to another
because the length of the run is measured using the variable time-step of
Equation~(\ref{timestep}).  However, on average we find that about $5 \times
10^6$ steps are needed for initial equilibration of the island shape, and we
run for about a further $10^7$ steps to accumulate reasonable statistics on
the ECS.  The average of the island shape over the run is performed by
taking 100 samples of the state of the lattice (one every $\sim 10^5$ steps),
and calculating the center of mass of the island.  From these samples we
then calculate the time-averaged occupation of the sites of the lattice over
the course of the run (being careful to take account of the variable
time-step defined in Equation~(\ref{timestep})) with the centers of mass
superimposed.  The ECS's shown in the figure are the 50 percent occupation
contour of the resulting distribution.

The value of $J$ for these runs is fixed at the $0.144{\rm eV}$ found in our
atom-embedding calculations for Cu, and we examine the variation of the
equilibrium crystal shape with $K$ for $T=300$K and $T=600$K.  The values
for $K$ from the atom-embedding calculations fall near the bottom of the
range shown in the figure (around $0.01{\rm eV}$), where the difference in
the equilibrium facet lengths of the islands is quite small.  As $K$ is
increased, the equilibrium length of the short facets decreases at a greater
and greater rate, until, at around $K=0.1$ they appear to vanish completely
and the island becomes triangular.  It is an open question whether there is
a phase transition in the system to the triangular state, and if there is,
whether it is a continuous transition.

In the experiments of Michely~{\it{}et~al.}\ a roughly constant ratio of the
lengths of the long and short sides of islands is observed, with a value of
$0.66\pm0.05$.  Assuming that the nearest-neighbor interactions between
adatoms in our simulations of Cu at 600K and the experiments on Pt at 700K
are comparable, we are able to extract a value of $K^\bigtriangleup = -K^\bigtriangledown \approx
-0.02{\rm eV}$ for the triangle interaction parameter for Pt.  The negative
sign here is indicative of the fact that long and short facets in Pt islands
are reversed with respect to the facets of the corresponding Cu islands.
The larger numerical value of the interaction corresponds to the greater
anisotropy of facet lengths seen in Pt by comparison with Cu.

 \section{Conclusions} We propose both a new model for the anisotropic
equilibrium faceting shapes seen in adsorbate islands on (111) surfaces, and
also a new non-local Monte Carlo algorithm for calculating ECS's in this or
any other conserved order-parameter model for faceting.  Using the
atom-embedding method of Finnis and Sinclair~\cite{finnis} we have calculated
the values of the parameters appearing in the model for Cu adsorbed on a
Cu~(111) surface, and find qualitative agreement with the island shapes seen
in the experiments of Michely and
co-workers~\cite{michely1,michely2,michely3}.

 \end{document}